# A NOVEL INTEGRATED APPROACH OF SOFTWARE DEVELOPMENT AND DESIGN PATTERN WITH X-CM

**Suprita Das, Rajesh Duvvuru and Sonal Raj**

*Department of Computer Science and Engineering*

*National Institute of Technology, Jamshedpur, Jharkhand, INDIA*

**ABSTRACT:**

*Expert evaluation in software life cycle is playing a vital role. In this paper, we have discussed about the formal framework for selecting the best model with some enhanced features. In our work, we have introduced a novel algorithm X-Chain Model (X-CM). Here, X-CM overcomes the limitation of V-model. In X-CM, we have concentrated on hierarchy of multiple independent sub-projects in project development phase. X-CM merges through the sub-projects binding them at their Integration Testing Phases. Hence combining binary modules at each stage into a new module through chains of "X" till the final level in the hierarchy is implemented. X-CM results in reducing the time and effort for software developer's. In addition to this, X-CM also rules out the dependency errors between the modules at each stage of combination. Here we have tested our proposed X-CM with Unit Testing and Integrating test. Finally, we proved that X-CM Model works better than Spiral, Prototype and V-Model.*

**Keywords:** X-chain; Sub-project; Divide and rule; Chain; Software lifecycle; Epitome

## [1] INTRODUCTION

Software Development Life Cycle (SDLC) is a series of identifiable steps that any software goes through in its lifetime. These steps have been uniquely identified to enhance the standard of the product to be developed. The X-CM which is being proposed by us is following similar steps with enhanced features like, efforts involved in large projects that comprise several sub-projects. X-CM is specially designed to handle the development of such projects. Existing software development models like Spiral model, V model and other SDLC model don't inherently provide development procedures for such projects and they are implemented by making some modifications in the original model but X-CM is specially designed to handle such projects. One appropriate example of such project may be the Apache's Hadoop [1] which is actually a collection of many projects like HDFS (Hadoop Distributed File System) [2] Hadoop Map reduce [3] and many others.

## [2] BACK GROUND WORK

For instance "online university management system". In this case the main system for online university management is first divided into many daughter systems as online college management systems, one system for each college under the university. Again each online college management system has further sub-projects like library management, academics.

In this project, we give emphasis on development of small subproject like library management system, academics, department then we merge it as X-CM pattern. It comprises series of 'X's that are finally merged to form the X-CM. In this model each individual 'X' uniquely corresponds to individual major subpart of a larger project. These subparts are supposed to be large enough and they themselves need separate stages of development involving intensive care and rigorous efforts for their completion. This model specially defines steps that individually pay attention to each and every steps of project development without leaving any scope, any error or fault that may lead to project failure at later stages of development or maintenance.

## [3] RELATED WORK

Earlier methodologies like Waterfall method, spiral model and V-model are called traditional software development methodologies and these are classified into the heavy weight methodologies [4] require defining and documenting a stable set of requirements at the beginning of a project There are mainly five phases which are characteristic of traditional software development method. The first step is to set up the requirements for the project and determine the length of time it will take to implement the various phases of development. Once the requirements are laid out, the next step moves into the design and coding where code is produced until the specific goals are reached. Development is often broken down into smaller tasks that are distributed among various teams based on skill. The testing phase comes after development phase to ensure issues are addressed early on. Then we go ahead for maintenance phase. Once the project nears completion and the developers are close to meeting the project requirements, the customer will become part of the testing and feedback cycle and the project was delivered after the customer satisfy with it.

## [3.1] THE EPITOME

The purpose of this paper is to describe a Epitome (A Pattern or Meta-model) which can be used to compare and contrast each of the above alternative life cycle models with each other and with constantly evolving user needs and finally to give emphasis on development of new model X-CM which overcome earlier problem.

For every application beyond the trivial, user needs is constantly evolving .Thus the system being constructed is always aim at a moving target. This is a primary reason for delayed schedules (caused by trying to make the software meet a new requirement it was not designed to meet) and software that fails to meet customer expectations (because the developers "froze" the requirements and failed to acknowledge the inevitable changes). For that purpose, for changing user needs we are giving focus on design and coding so that it will solve user problem. The earlier software development methods are dependent on a set of predetermined processes and on-going documentation which is written as the work progresses and guides further development [5]. The success of a project which is approached in this way relies on knowing all of the requirements before development begin and means that implementing change during the development life cycle can be somewhat problematic. To overcome such type of problem we are introducing new model known as : X-chain model. In X-chain development, rather than a single large process model that implemented in traditional SDLC, the development life cycle is divided into smaller parts, called basically incremental step in which each of these increments touches on each of the traditional phases of development and similar way we are performing testing to make the software rigid and effective X-chain software development methods basically developed to provide more customer satisfaction, to reduce bug rates, and to accommodate changing business requirement during development and coding phases [6].

Below figure describes how user needs acquire over time. It is recognized that the different function shown is neither linear nor continuous in reality. The units on the x-axis can be either months or years), but could be assumed to be non-uniform, containing areas of compression and decompression and the units of the scale on the y-axis assumed to be some measure of the amount of functionality (such as DeMarco's "Bangs for the Buck") [7].

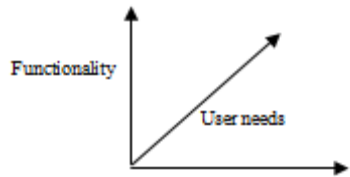

**Figure 1.Constantly acquiring user needs**

# [4] RELETED WORK

## [4.1] ASSUMPTIONS AND NOTATIONS

In X-CM individual firstly subprojects are treated as 'X' where each X has a series of precisely defined steps that treat these subprojects as major projects. Each 'X' has been divided into two parts: Upper X and lower X. The upper X consists of major development phase while lower X consists of complementary steps that compliment the upper corresponding step. For example the requirement specification and feasibility study in the upper 'X' is major part which is being complemented by customer review in the lower 'X' of the 'X-CM which means, when the feasibility study is completed and the software requirement specification (SRS) document has been produced, before transiting to the next phase user requirement documents will be reviewed by the customer. Similarly all the major processes are being complemented by subsidiary processes present in the lower half of 'X'. Such complementary processes eliminate the possibility of errors or fault being overlooked. One by one the development process transits from one phase to another from left to right i.e. from feasibility study to integration and integrity check through system architecture design, coding and unit testing.

Since X-CM is specially designed for the development of larger projects, parallel process for testing is provided for each phase of the model to eliminate errors. Starting from requirement analysis and feasibility study, it is supported by customer review to minimize ambiguities and inconsistency in the development of SRS document. It also helps to understand the user's requirement in a better and complete manner. In this phase main system is divided into modules. This phase is complemented by component integrity check which ensures that the modules developed are optimized in both number and size. Module design phase is complemented by module integrity check which ensures the modular integrity and confirm that modules are having high cohesion and less coupling. Coding is complemented by unit testing to remove any error at the module level. Integration and system testing is complemented by system architectural check which matches the integrated system with logical architectural design of the system developed in 'system architecture design' phase. This phase also act as the linking place for different modules to get merged. User acceptance testing complements the maintenance phase which finds further scope of improvement of features by beta testing and customer preview of the software [3][4]. User review and testing processes have been given prime importance for each individual phase as shown in the figure. Finally the individual projects are merged in the integration phase after which the integration testing is carried out .The phases described inside 'X' diagram are common for each subproject and repeated for each subproject individually. But the 'Maintenance and Customer Preview' and 'User acceptance' are the processes that will be performed after the completion of merging process and only once that's why they have been kept outside 'X'.

## [4.2] THE X-CM

In this model we have followed the approach of Divide-& Conquer. The various phases involved in individual 'X-chain are as follows. The reason for selecting the X-Design to represent the Software Development Life Cycle [1] takes into consideration the following two reasons,

- Firstly, the X-CM is divided vertically, in the form of "Phase Design" and "Implementation & Deployment Phase". Where Phase design is kept in the left part of X-CM model and in right part of X-CM, implement and deployment phases are set aside.

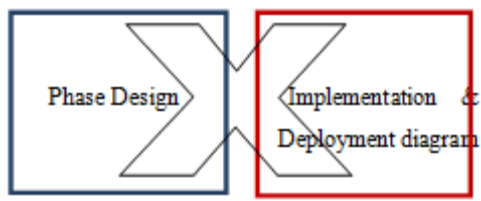 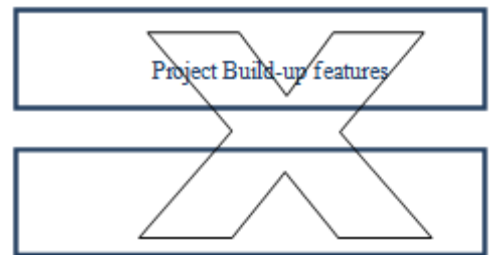

**Figure 1 Prototype of X-CM (Vertical Bifurcation)**     **Figure 2 Prototype of X-CM**

- Secondly, X-CM model is bifurcated horizontally. The upper portion of the X is concerned with the prominent features that will lead to the completion of the software development. Whereas the lower part deals with the corresponding check or test processes. In which will ensure that the progress made in the top phases are completely error free.

## [4.3] REQUIREMENT ANALYSIS AND FEASIBILITY STUDY-CUSTOMER REVIEW

Requirement analysis and feasibility study is the first phase in the X-chain software development process model. In this phase the requirements of the proposed software system are gathered by collecting and analyzing the needs of the users. The main aim of feasibility study is to determine whether it would be financially and technically feasible to develop the product. In this phase, the process of requirement analysis and feasibility study can be better implemented by collecting the customers review. The SRS document is prepared which typically describes the systems functional, interface, performance, data, security, etc. requirements as expected by the users.

This phase is complemented by the customer review which ensures to get complete user's requirements

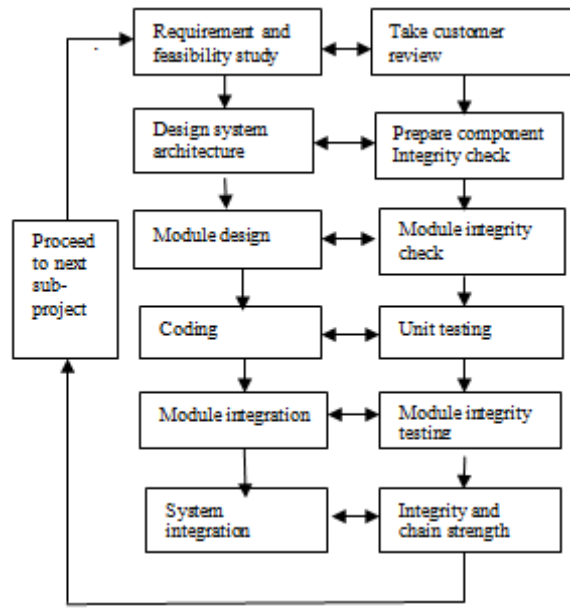

Figure 3. Flow Diagram for X-CM

### [4.4] DESIGN SYSTEM ARCHITECTURE/ COMPONENT INTEGRITY CHECK

In this phase we design system architecture for each module and in parallel we go for integration check.

### [4.5] MODULE DESIGN / MODULE INTEGRITY CHECK

In this phase each module is designed. The module design phase can also be referred to as low-level design. The designed system is broken up into smaller units or modules and each of them is explained so that the programmer can start coding directly. The low level design document or program specifications will contain a detailed functional logic of the module. The sub-process of this phase is 'module integration check' which does the same job to the module design which 'component integrity check' does for the 'system architectural design' process. It checks for the architectural integrity of the module.

### [4.6] CODING / UNIT TESTING

The purpose of the coding and unit testing phase (also called the implementation phase) of software development is to translate the software design into source code. Each component of the design is implemented as a program module. The end-product of this phase is a set of program modules that have been individually tested by Unit testing process. During this phase, each module is unit tested to determine the correct working of all the individual modules. It involves testing each module in isolation as this is the most efficient way to debug the errors identified at this stage. Unit testing is a method by which individual units of source code are tested to determine if they are fit for use. A unit is the smallest testable part of an application. In procedural programming a unit may be an individual function or procedure. Unit tests are created by programmers or occasionally by white box testers

The purpose is to verify the internal logic code by testing every possible branch within the function, also known as test coverage. Static analysis tools are used to facilitate in this process, where variations of input data are passed to the function to test every possible case of execution.

## [4.7] MODULE INTEGRATION AND TESTING

In the X-CM this is the one of the most important phase where the integration of the subprojects takes place. Until now the subprojects were being treated like individual entities and were being developed individually, but it's now time for merging the segregated projects one by one to form the complete project.

A subproject has crossed the coding and testing phase and all the modules related to the subproject are ready to be merged. In this phase individual modules are integrated and tested one by one which is module integrity check. Later when a subproject is complete with all its modules merged together and tested, it's time for merging it with already completed subproject. Let's see how it is done. The two subprojects have got all their individual modules integrated and tested. Talking in terms of 'X', the modules integration part is complete. Now the two 'X's are merged together with their merging junction, at integration and integrity testing at the upper 'X' and Chain strength check at the lower 'X'. The integration is done in terms of interfacing the two subprojects. The level of integration or the strength of integration is dependent on project concerned and the development team. After the integration is completed the two individual 'X' are merged into one as shown in figure .After the merging is completed its time for the complementary process - Chain strength check.

Chain strength is the measure of integral strength of a module with the system with respect to the system architectural design. It is measured in integrating testing phase. The value of chain strength ranges from 0 to 1. Value of chain strength depends on how the integration of modules is similar to the logical architecture of the system designed in 'system architecture design' phase. Chain strength is measured after integrating each module by the developers. If module integration is fully compatible with the system i.e. it completely matches with the designed system architecture then its chain strength value will be 1. If system doesn't works as supposed after integration and it completely doesn't match with the desired architecture of system then the chain strength for that module integration will be 0. Chain strength of each module integration is measured and after the complete integration, overall integration quality can be measured by calculating the average value of all chain strength values measured so far.

Chain strength provides the measuring parameter for the integration quality of system. Value of chain strength of a system measures the integral quality of that system which can be used for future references. A minimum value of chain strength can be set by developers for the system to achieve for considering the system of high quality. If developers find the system having enough chain strength hence they can proceed with next phase of software development otherwise modifications can be done in modules and integration.

### [4.8] SYSTEM MAINTENANCE /USER ACCEPTANCE TESTING

In the Maintenance phase those errors are rectified which were not apparently found in the product development phase. This phase is also responsible in improving and enhancing the system functionality in accordance with the customer's requirement. Maintenance process works parallel and in accordance with the user acceptance testing process.

## [5] TEST AND RESULTS

### [5.1] UNIT TEST ON X-CM

Here we have performed Unit Testing on our proposed X-CM model. We have observed that the X-CM model has given an efficient way of unit test compared with Spiral, V-Model and Prototype [8][9].

### [5.2] INTEGRATION TEST ON X-CM

Next, we applied Integration test on X-CM model. Here, we have combined different sub X-CM model to form a complete X-CM model. Thus, the Integration Testing acts as the combinatory element for the current and the previous sub-projects [2]. This has the following advantages in its part –The Integration testing at each stage leads to better cohesion of the combined or merged unit formed for the testing purpose. Figure 4, describes clearly the integration mechanism. After integration we have we observed that X-CM performs better than V-Model, Spiral and Prototype [9][10][11].

It also ensures that the two different units ( of sub-projects) being considered are free from errors after combination and all dependency problems between them are resolved This method also eliminates the need for a separate system testing procedure since the final Integration test will suffice to test the Project as a whole. Also, by conducting integration test at each step, the load of a single integration test is greatly reduced and distributed [12][13]. We can apply X-CM model in larger projects ERP Systems [14] and Defense applications etc.

## [6] CONCLUSION

The X-CM leads to better automation of the Software Development process in general and large hierarchical projects in particular. It also reduces the effort of testing and also eliminates the need for separate System testing. The Divide and Conquer approach leaves very less scope for errors to be retained after the completion of the project development. Moreover, its perfection is also sought by the rigorous step-wise testing. It is a very useful methodology to be adopted in the modern software development process to replace the traditional heavyweight development life cycle. Hence, the X-CM possesses the ideal design criteria for every Software Development.